\documentclass[conference,letterpaper]{cssconf} 

\IEEEoverridecommandlockouts                              
\usepackage{epsfig}        
\usepackage{mathptmx}      
\usepackage{amsmath}       
\usepackage{amssymb}       
\usepackage{siunitx}
\usepackage[english]{babel}
\usepackage{csquotes}
\usepackage{hyperref}
\usepackage{epstopdf}
\usepackage{graphicx}

\usepackage{booktabs}      
\usepackage{microtype}     
\usepackage[
  backend=biber,
  style=ieee,
  doi=false,
  url=false,
  isbn=false,
  date=year
]{biblatex}

\title{\LARGE \bf Control-Oriented Modeling of Heat Exchangers with Improved Low-Flow Dynamics}

\author{Noah Hofmann$^{1}$, Stefan L. H\"olzl$^{1}$, Robert Bauer$^{2}$ and Martin Steinberger$^{1}$
\thanks{$^{1}$ Christian Doppler Laboratory for Optimized Model-Based Operation of Interconnected Systems, Institute of Automation and Control, Graz University of Technology, Graz, Austria \texttt{noah.hofmann@tugraz.at}
\protect\\
The financial support by the Austrian Federal Ministry of Economy, Energy and Tourism, the National Foundation for Research, Technology and Development 
and the Christian Doppler Research Association is gratefully acknowledged.}
\thanks{$^{2}$ Kristl, Seibt \& Co. Gesellschaft m.b.H., 8052 Graz, Austria}
}

\newcommand{\myFigure}[5]{
    \begin{figure}[ht]
    \centering
    \includegraphics[keepaspectratio,#2]{#1}
    \caption[#4]{#3}
    \label{fig:#5}
    \end{figure}
}

\renewcommand{\vec}[1]{\mathbf{#1}} 
\newcommand{\gvec}[1]{\pmb{#1}}

\begin{document} 
    \maketitle 
    \begin{abstract} 
        Existing low-order models of heat exchangers can lose dynamic accuracy at low flow rates, when fluid residence times become significant. 
        This paper proposes two control-oriented models that improve the low-flow performance. The first extends a finite-volume formulation with the first spatial temperature moment, 
        whereas the second uses Chebyshev--Lobatto collocation. Both capture the coupled heat-transfer and transport dynamics and improve the performance when the system behavior is dominated by the inlet-temperature dynamics.
        At equal state dimension, comparisons with a high-order reference show more accurate delayed outlet-temperature
        responses while retaining steady-state accuracy.
    \end{abstract}
    \section{Introduction} 
        \label{sc:Introduction}
        Heat exchangers are used in a multitude of applications, ranging from district heating networks and domestic heating \cite{klein2026cfd} to industrial production \cite{horvathova2023real} to automotive applications \cite{buck2026neural} to conditioning of process media in research and development \cite{zengReviewHeatTransfer2020}.
        The latter typically requires accurate temperature control to maintain prescribed thermal operating conditions \cite{zengReviewHeatTransfer2020}.
        Model-based control is promising for meeting these requirements \cite{horvathova2023real}, \cite{oravec2017advanced}, \cite{michelModelBasedControl2014}, but it depends on the complexity and quality of the underlying model.
        Here, a dilemma arises: accurate models are often of high order---e.g. infinite-dimensional---\cite{maidiBoundaryGeometricControl2009}, \cite{christofidesNonlinearRobustControl2001}, whereas the low-order models are often inaccurate---especially under low-flow conditions---and/or do not permit physical interpretation \cite{buck2026neural}, \cite{oravec2017advanced}, \cite{michelAccurateLoworderDynamic2013}.
        To solve this dilemma, the present paper proposes two \emph{control-oriented, low-order models of heat exchangers that are more accurate even under low-flow conditions}, thus providing a sound basis for model-based estimation and control.

        An ideal control-oriented heat exchanger model describes the dynamics from the inlet to the outlet temperature on both sides, while explicitly accounting for the strong flow rate dependency of these dynamics.
        Distributed-parameter models, based on physical insight, satisfy this requirement, but their direct use in real-time algorithms is limited by their infinite-dimensional nature \cite{christofidesNonlinearRobustControl2001,maidiBoundaryGeometricControl2009}
        and conventional upwind discretization schemes---like the \emph{method of lines (MOL)}---introduce numerical diffusion \cite{hundsdorferNumericalSolutionTimeDependent2003}.
        In \cite{michelAccurateLoworderDynamic2013}, a distributed-parameter model was discretized using the finite-volume method, with a low number of cells, and the steady-state accuracy was improved by incorporating the stationary solution of the underlying PDE; thus, a low-order model with high steady-state accuracy was obtained.
        This so-called \emph{power-balance approximation} was used for model-based control in \cite{michelModelBasedControl2014}.

        The main problem with the power-balance (PB) approximation is that it relies on a small fluid residence time, hence it becomes inaccurate for low flow rates (low-load operation), and even shows unphysical behavior for rapid changes of the inlet temperatures---which may be the case if a 3-way valve is used to bypass the heat exchanger in low-load operation.
        It is shown in this paper that in the pure-transport case, i.e. no heat transfer, the single-volume power-balance approximation corresponds to a $\left(1/1\right)$-Padé approximation---which shows a step-wise change to $-1$ at the output under a step-wise change of the input to $+1$, and thus a step-wise change of the outlet temperature \emph{in the wrong direction}!

        \emph{This paper proposes} two finite-dimensional approaches for improving the approximation of convective transport without introducing explicit time-delay operators.
        The two approaches are similar to the power-balance approximation both in terms of complexity and steady-state accuracy, yet with \emph{improved low-flow performance}.
        The \emph{first approach} extends the power-balance approximation by retaining the first spatial temperature moment in addition to the mean temperature of each finite volume. 
        This additional state captures the leading-order spatial asymmetry of the local temperature field and thereby improves the representation of convective transport when the fluid residence time is significant, yet without requiring a finer finite-volume discretization.
        The \emph{second approach} uses Lagrange interpolation at Chebyshev--Lobatto (CL) nodes to approximate the spatial derivative through a differentation matrix acting on the vector of all nodal temperatures. 
        The resulting approximation can achieve high spatial accuracy with comparatively few nodes.

        The proposed models (presented in Sections~\ref{sc:First_Moment_Power_Balance_Extension} and~\ref{sc:Chebyshev_Lobatto_Collocation}) are evaluated against the original power-balance approximation and a high-order method-of-lines reference model (both summarized in Section~\ref{sc:Preliminaries}) at comparable state dimensions, with particular emphasis on low-flow operation.
        The numerical evaluation is given in Section~\ref{sc:Results and Comparison}, followed by concluding remarks in Section~\ref{sc:Conclusion}.

    \section{Preliminaries}
        \label{sc:Preliminaries}
        Consider a countercurrent heat exchanger with two fluid circuits, as shown in \autoref{fig:schematic_heat_exchanger_}, where $I$ and $II$ denote the cold and hot sides, respectively. 
        The position along the effective heat-exchanger length $L$
        is denoted by $x \in \left[0,L\right]$.
        \myFigure{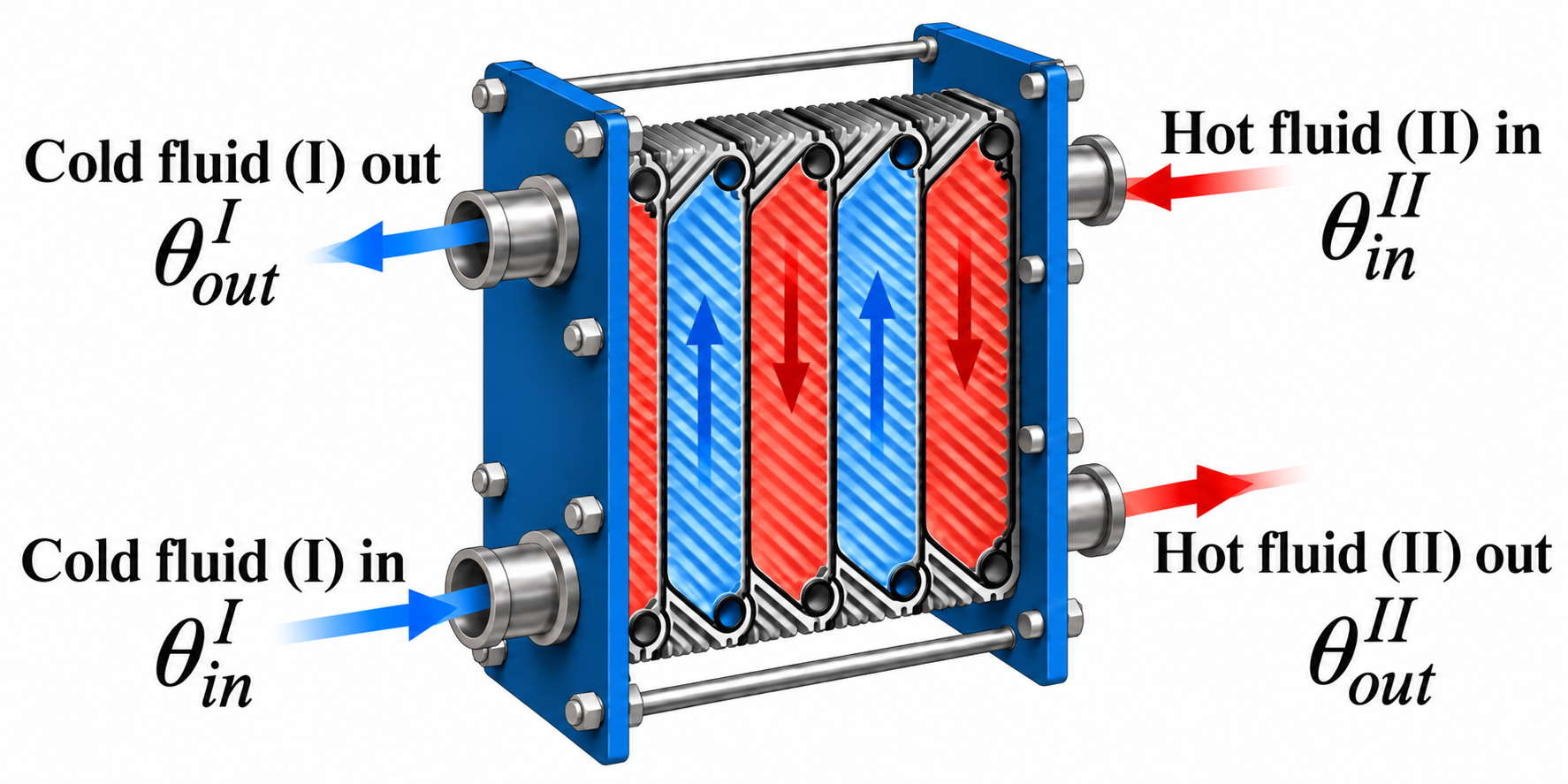}
        {width=0.5\columnwidth}
        {Schematic representation of a heat exchanger. Image generated by \texttt{ChatGPT Images 2.5}}
        {blalbalba}
        {schematic_heat_exchanger_}
        \subsection{Distributed-Parameter Model}
        \label{ss:Distributed_parameter_model}
        A distributed-parameter model of the heat exchanger can be derived by applying an energy balance under the following assumptions: \emph{incompressible fluids}, \emph{one-dimensional plug flow},
        \emph{negligible heat conduction of the plates in the direction of the fluid flow}.
        Under these constraints, a simplified model can be formulated as a set of two hyperbolic PDEs \cite{maidiBoundaryGeometricControl2009}: 
        \begin{subequations}
            \label{eq:Heat_PDEs}
            \begin{align}
                \frac{\partial \theta^I}{\partial t} + v^I \frac{\partial \theta^I}{\partial x}+\alpha^I U (\theta^I - \theta^{II}) &= 0,\label{eq:heat_PDE_1}\\
                \frac{\partial \theta^{II}}{\partial t} + v^{II} \frac{\partial \theta^{II}}{\partial x}+\alpha^{II} U (\theta^{II} - \theta^{I}) &= 0. \label{eq:heat_PDE_2}
            \end{align}
        \end{subequations}
        Here, $\theta^{I}(x,t)$ and $\theta^{II}(x,t)$ denote the temperature fields, $v^I~>~0$ and $v^{II}~<~0$ are the signed flow
        velocities for countercurrent operation, and $U$ is the overall heat-transfer coefficient.
        The coefficients $\alpha^{I}$ and $\alpha^{II}$ account for both geometric and material properties of the respective circuits. 
        The boundary conditions are imposed at opposite ends due to the countercurrent flow: \mbox{$\theta^{I}(0,t) = \theta^{I}_{in}(t)$}, \mbox{$\theta^{II}(L,t) = \theta^{II}_{in}(t)$.}

        In general, the overall heat-transfer coefficient $U$ depends on both flow rates $q_i$ $(i \in \{I,II\})$,
        where $q_i = |v^i| A_i$, $A_i$ denotes the effective channel cross-sectional area, and $v^i$ represents the mean channel velocity.
        The overall heat transfer coefficient $U$ can be determined from a Nusselt correlation, see \cite{michelAccurateLoworderDynamic2013}.
        \subsection{Baseline Power Balance Model}
        \label{ss:Baseline_Power_Balance_model}
        To obtain a finite-dimensional approximation of the distributed-parameter model, the power balance scheme proposed in \cite{michelAccurateLoworderDynamic2013} is employed. 
        The key idea of this method is to derive the stationary temperature profiles of the heat exchanger from the underlying PDE model given in \eqref{eq:Heat_PDEs}. 
        These profiles are subsequently used to establish algebraic relationships between the volume-averaged temperatures and the corresponding outlet temperatures. 
        In this way, the reduced-order model preserves the stationary energy balance and achieves steady-state accuracy even when only a single finite-volume element is used.
        \subsection{Pure Transport Analysis}
        \label{ss:Pure_transport_analysis_PB}
        To investigate the approximation behavior of the PB scheme under pure transport conditions, heat transfer is neglected by assuming an overall heat transfer coefficient
        of $U = 0$ and considering only one volume element. In this specific scenario, the two fluid circuits are fully decoupled; therefore only circuit $I$ is analyzed. 
        For this special case the transfer function between inlet and outlet temperature of circuit $I$ is given by
        \begin{equation}
            \label{eq:Dead_Time_Approximation_PB}
            G_{I}(s) = \frac{\theta^{I}_{out}(s)}{\theta^{I}_{in}(s)} = \frac{-s+\frac{2}{\tau_{I}}}{s+\frac{2}{\tau_{I}}} = \frac{1-\frac{\tau_{I}}{2} s}{1+\frac{\tau_{I}}{2}s} \approx e^{-s \tau_I}.
        \end{equation}
        Here, $\tau_{I} = \frac{V_{I}}{|q_{I}|}$, where $V_{I}$ is the fluid volume of the cold side circuit inside the heat exchanger.
        In the pure-transport case, the single-volume PB model yields the first-order diagonal Padé approximation of the transport delay \cite{bakerPadeApproximantsSecond1996}. 
        The approximation has direct feedthrough and a right-half-plane zero at $s_1 = \frac{2}{\tau_I}$, resulting in an instantaneous inverse outlet response to an inlet-temperature step.
        To eliminate the direct inlet-to-outlet feedthrough while improving the short-time transport response, the power balance approximation is extended by retaining the first spatial moment.

    \section{First-Moment Power Balance Extension}
        \label{sc:First_Moment_Power_Balance_Extension}
        The proposed extension introduces a second state variable, the first spatial moment, for each volume element.
        The heat exchanger of effective length $L$ is divided into $M$ elements of equal length $\Delta l = {L}/{M}.$
        For the $j$-th element, $\gamma \in \left[l_{j-1}, l_j\right]$ is introduced, where \mbox{$l_j = l_{j-1}+ \Delta l$},  \mbox{$j = 1,2,...,M$.}
        By defining the local coordinate, $\eta = \frac{\gamma - l_{j-1}}{\Delta l} \in \left[0,1\right]$,
        $\gamma$ can be expressed as $\gamma = l_{j-1} + \eta \Delta l$, with its partial derivative $\frac{\partial \eta}{\partial \gamma} = \frac{1}{\Delta l}$.
        With this, the infinite-dimensional model of the heat exchanger \eqref{eq:Heat_PDEs} can be partitioned into $M$ volume elements, resulting in two local PDEs per element:
        \begin{subequations}
            \label{eq:discretized_heat_PDEs}
        \begin{align}
            \frac{\partial \theta_j^I}{\partial t} = & -\frac{v^{I}}{\Delta l} \cdot \frac{\partial \theta_j^I}{\partial \eta} - \alpha_I U (\theta_j^I - \theta_j^{II}),\\
            \frac{\partial \theta_j^{II}}{\partial t} = & -\frac{v^{II}}{\Delta l} \cdot \frac{\partial \theta_j^{II}}{\partial \eta} - \alpha_{II} U (\theta_j^{II} - \theta_j^{I}).
        \end{align}
        \end{subequations}
        \subsection{Definition of Spatial Moments}
        \label{ss:Definition_of_spatial_moments}
        For each circuit $i \in \{I, II\}$ and element $j$, the mean temperature and the first spatial moment of the $j$-th cell
        \begin{equation}
            \label{eq:mean_temperature_FSM_definition}
            \bar\Theta_j^i = \int_0^1 \omega_0(\eta) \theta_j^i(\eta, t) d\eta, \quad m_j^i = \int_{0}^{1} \omega_1(\eta) \theta_j^i(\eta, t) d\eta, 
        \end{equation}
        are defined using the weighting functions
        $\omega_0(\eta) = 1$ and  $\omega_1(\eta) = 2\eta - 1$, respectively.
        The first spatial moment associated with the second weighting function $\omega_1(\eta)$ captures the leading-order spatial asymmetry of the temperature profile and distinguishes profiles that increase or decrease along the local coordinate.
        The weighting function varies linearly from $-1$ at $\eta=0$ to $1$ at $\eta=1$ and has zero mean.
        \subsection{Model Formulation}
        \label{ss:Model_formulation}
        Taking the temporal derivatives of the moments in \eqref{eq:mean_temperature_FSM_definition}, substituting in the local PDEs \eqref{eq:discretized_heat_PDEs},
        and integrating the convective terms by parts yields, for $j = 1,2, \ldots, M$,
        \begin{align}
            \label{eq:ODEs_first_spatial_moment}
            \frac{d\bar\Theta_j^I}{dt} &= -\frac{v^{I}}{\Delta l} \cdot \left(\theta_j^I - \theta_{j-1}^I\right)- \alpha^I U (\bar\Theta_j^I - \bar\Theta_j^{II}),\\
            \frac{d\bar\Theta_j^{II}}{dt} &= -\frac{v^{II}}{\Delta l } \cdot \left(\theta_j^{II} - \theta_{j-1}^{II}\right)- \alpha^{II} U (\bar\Theta_j^{II} - \bar\Theta_j^{I}),\nonumber\\
            \frac{dm_j^I}{dt} &= -\frac{v^{I}}{\Delta l} \left(\theta_j^I(t) + \theta_{j-1}^I(t) - 2 \bar\Theta_j^I\right) - \alpha^I U \left(m_j^I - m_j^{II}\right), \nonumber\\
            \frac{dm_j^{II}}{dt} &= -\frac{v^{II}}{\Delta l} \left(\theta_j^{II}(t) + \theta_{j-1}^{II}(t) - 2 \bar\Theta_j^{II}\right) - \alpha^{II} U \left(m_j^{II} - m_j^{I}\right)\nonumber.
        \end{align}

        Here, $\theta_{j-1}^{i}$ and $\theta_{j}^{i}$ denote the temperature at the left and right boundaries of element $j$, respectively.
        The first two equations describe the time dependence of the mean temperatures, whereas the last two govern the corresponding first spatial moments.
        The convective terms in \eqref{eq:ODEs_first_spatial_moment}, i.e. the terms proportional to $v^{i}$, depend on the left and right boundaries of each element. 
        These boundary temperatures can be reconstructed from the mean temperature $\bar{\Theta}_j^i$ and the first spatial moment $m_j^i$ by prescribing the shape of the local temperature profile. 
        Following the original PB approach, this shape is chosen according to the stationary solution of the distributed-parameter model \eqref{eq:Heat_PDEs}.
        Accordingly, the temperature profile within element $j$ is written as 
        \begin{equation} 
            \label{eq:stationary_profile}
            \theta_j^i(\eta,t) = \theta_{j-1}^i(t) + \bigl(\theta_j^i(t)-\theta_{j-1}^i(t)\bigr) \Phi_\Gamma(\eta),
        \end{equation}
        where 
        \begin{equation}
            \label{eq:Phi_Gamma_eta}
            \Phi_\Gamma(\eta) =  \frac{1-e^{-\Gamma \eta}} {1-e^{-\Gamma}}, \quad \Gamma \neq 0,
        \end{equation}
        and the dimensionless profile parameter is $\Gamma = U \Delta l\left(\frac{\alpha_{I}}{v^I} + \frac{\alpha_{II}}{v^{II}} \right).$
        Substitution of \eqref{eq:stationary_profile} into the mean temperature and moment definitions \eqref{eq:mean_temperature_FSM_definition} gives
        \begin{subequations}
        \label{eq:stationary_profile_reconstruction}
        \begin{align}
            \label{eq:mean_temperature_boundary}
            \bar{\Theta}_j^i &= \theta_{j-1}^i + \xi(\Gamma) \left(\theta_{j}^i-\theta_{j-1}^i\right),\\
            \label{eq:spatial_moment_boundary}
            m_j^i &= \mu(\Gamma) \left(\theta_{j}^i-\theta_{j-1}^i\right),
        \end{align}
        where
        \begin{equation}
            \label{eq:xi_mu}
            \xi(\Gamma) = \int_0^1 \Phi_\Gamma(\eta)\,d\eta, \quad \mu(\Gamma) = \int_0^1 (2\eta-1)\Phi_\Gamma(\eta)\,d\eta. 
        \end{equation}
        \end{subequations}
        The coefficient $\xi(\Gamma)$ relates the element mean to its boundary temperatures, whereas $\mu(\Gamma)$ relates the first spatial moment to 
        the temperature difference across the element.
        In particular, \eqref{eq:spatial_moment_boundary} gives $\theta_j^i-\theta_{j-1}^i = m_j^i/\mu(\Gamma)$.
        Thus, the first spatial moment determines the reconstructed temperature variation across the element, including its sign, while the mean temperature determines the overall temperature level.
        Solving \eqref{eq:mean_temperature_boundary} and \eqref{eq:spatial_moment_boundary} for the two boundary temperatures gives:
        \begin{subequations}
            \label{eq:interface_conditions_spatial_moment}
            \begin{equation}
                \theta_{j,L}^i = \bar{\Theta}_j^i - \frac{\xi(\Gamma)}{\mu(\Gamma)}m_j^i, \quad  \theta_{j,R}^i = \bar{\Theta}_j^i + \frac{1-\xi(\Gamma)}{\mu(\Gamma)}m_j^i.
            \end{equation}
            Here, $\theta_{j,L}^i$ and $\theta_{j,R}^i$ denote the temperatures reconstructed from the state variables of element $j$ at $\eta=0$ and $\eta=1$, respectively.
            Because the reconstruction can be performed individually for every element, the right boundary temperature $\theta_{k,R}^i$ of element $k$ and the left boundary temperature $\theta_{k+1,L}^i$
            of element $k+1$ are not necessarily identical, in general.
            A unique temperature must nevertheless be assigned to each interface before the boundary terms in \eqref{eq:ODEs_first_spatial_moment} can be evaluated. 
            To resolve this ambiguity, a unique boundary temperature is selected according to the flow direction in the respective circuit.
            For positive velocity $v^i>0$, information is transported from left to right. The temperature at an internal interface $k$ is therefore taken as the right boundary temperature of element $k$.
            For negative velocity $v^i<0$, information is transported from right to left, and the interface temperature is taken as the left temperature of element $k+1$. 
            At the external inlet boundary, the prescribed inlet temperature is used. The resulting upwind boundary temperature is
            \begin{equation}
                \label{eq:upwind_boundary_conditions}
                \hat{\theta}^{i}_k = \begin{cases}
                    \theta^{i}_{\mathrm{in}}, & k = 0, \hspace{2.17cm} v^{i} > 0,\\
                    \theta^{i}_{k,R}, & k = 1,2,\ldots, M, \hspace{0.85cm}  v^{i} > 0,\\
                    \theta^{i}_{k+1,L}, & k = 0,1, \ldots, M-1, \hspace{0.3cm}  v^{i} < 0,\\
                    \theta^{i}_{\mathrm{in}}, & k = M, \hspace{2cm}  v^{i} < 0.
                \end{cases}
            \end{equation}
        \end{subequations}
        Accordingly, the left and right boundary temperatures of element $j$ are replaced by $\hat{\theta}^{i}_{j-1}$ and $\hat{\theta}^{i}_j$, respectively.
        \subsection{Pure Transport Analysis}
        \label{ss:Pure_transport_analysis_spatial_moments}
        When heat transfer is again neglected, i.e. $U \to 0$, it follows that $\Gamma \to 0$. The profile function \eqref{eq:Phi_Gamma_eta} satisfies \mbox{$\lim_{\Gamma \to 0} \Phi_{\Gamma}(\eta) = \eta$},
        the coefficients $\xi$ and $\mu$ reduce to the constant values: $\xi(0) = 1/2$ and $\mu(0) = 1/6$, respectively.

        Using the same procedure as in \autoref{ss:Pure_transport_analysis_PB}, the transport phenomenon in the decoupled case of the two circuits can be analyzed. In the following, only circuit $I$
        is considered, circuit $II$ exhibits equivalent behavior in the decoupled case.
        Setting $U=0$ in \eqref{eq:ODEs_first_spatial_moment}, applying the boundary conditions \eqref{eq:interface_conditions_spatial_moment}, and substituting the limiting values of $\xi$ and $\mu$ yield,
        for circuit $I$,
        \begin{align}
            \label{eq:Dead_time_Approximation_Spatial_Moments}
            G_I(s) &= \frac{\theta_{\mathrm{out}}^I (s)}{\theta_{\mathrm{in}}^I (s)} =\frac{1 - \frac{1}{3} \tau_I s}{1+ \frac{2}{3} \tau_I s + \frac{1}{6} \tau_I^2 s^2} \approx e^{-s \tau_I}.
        \end{align}
        Thus, when only one volume element with two state variables is used, the pure-transport delay is represented exactly by the $\left(1/2\right)$-Padé approximation.
    \section{Chebyshev--Lobatto Collocation}
        \label{sc:Chebyshev_Lobatto_Collocation}
        As a second approach, the distributed-parameter model is approximated by orthogonal collocation on Chebyshev--Lobatto (CL) nodes \cite{trefethenSpectralMethodsMATLAB2000}, \cite{weidemanMATLABDifferentiationMatrix2000}, \cite{binousChebyshevOrthogonalCollocation2015}.
        In contrast to finite-volume methods, which approximate the convective derivative using local differences between adjacent volume elements,
        the collocation method represents the temperature fields by global Lagrange interpolation polynomials. 
        The spatial derivatives at the collocation nodes can therefore be expressed as a matrix-vector product involving the nodal temperatures.
        Since the resulting differentiation matrix is generally dense, the derivative at each node depends on the temperatures at all nodes. 
        This global spatial approximation can provide high accuracy with comparatively few degrees of freedom for sufficiently smooth temperature profiles.
        \subsection{Collocation Nodes and Boundary Conditions}
        \label{ss:Collocation_Nodes_Boundary_Conditions}
        The spatial coordinate is normalized according to $\zeta=x/L\in[0,1]$ and subsequently mapped to $z=2 \zeta-1\in[-1,1]$. 
        For $N$ collocation nodes, the CL points are defined as 
        \begin{equation} 
            \label{eq:Chebyshev_Lobatto_nodes} 
            z_n=-\cos\left(\frac{n\pi}{N-1}\right), \qquad n=0,1,\ldots,N-1. 
        \end{equation} 
        The resulting nonuniform node distribution mitigates Runge's phenomenon, which refers to large oscillations near the interval boundaries that may arise when using equally spaced interpolation nodes
        \cite{fornbergRungePhenomenonSpatially2007}.
        The corresponding physical positions are $\zeta_n=(z_n+1)/2$. The nodes include both boundaries and are concentrated near the ends of the spatial domain.
        For the countercurrent configuration, the inlet boundary conditions are imposed at opposite boundaries, \mbox{$\theta^I(z_0,t)=\theta_{\mathrm{in}}^{I}(t)$}, \mbox{$\theta^{II}(z_{N-1},t)=\theta_{\mathrm{in}}^{II}(t)$.}
        Consequently, the differential equations of side~$I$ are evaluated at nodes $n=1,2,\ldots,N-1$, whereas those of side~$II$ are evaluated at nodes $n=0,1,\ldots,N-2$. 
        The outlet temperatures are given by \mbox{$\theta_{\mathrm{out}}^{I}=\theta^I(z_{N-1},t)$}, \mbox{$\theta_{\mathrm{out}}^{II}=\theta^{II}(z_0,t)$.}
        For each circuit $i \in \{I, II\}$, the nodal temperatures are defined as $\theta^{i}_n(t) = \theta^{i}(z_n,t)$, $n= 0,1,\ldots, N-1,$
        and collected in the full nodal-temperature vector $\gvec{\theta}^{i}(t) = \begin{bmatrix} \theta^{i}_0 & \theta^{i}_1 & \ldots & \theta^{i}_{N-1} \end{bmatrix}^T \in  \mathbb{R}^N$.

        \subsection{Lagrange Interpolation and Differentiation Matrix}
        \label{ss:Lagrange_interpolation_and_differentiation_matrix}
        For each circuit $i\in\{I,II\}$, the temperature field is approximated by a unique polynomial of degree at most $N-1$ that interpolates the nodal temperatures:
        \begin{equation} 
            \label{eq:Lagrange_temperature_approximation}
            \theta^{i}(z,t) \approx \sum_{n=0}^{N-1}\theta^i_n(t)L_n(z), 
        \end{equation} 
        where the $n$-th Lagrange basis polynomial is defined as $L_n(z)=\prod_{k=0,k\neq n}^{N-1}(z-z_k)/(z_n-z_k)$.
        The basis functions satisfy the cardinality condition $L_n(z_m)=\delta_{nm}$, with $\delta_{nm} = 1$ for $n=m$ and zero everywhere else, such that the approximation 
        in \eqref{eq:Lagrange_temperature_approximation} exactly reproduces the nodal temperatures, $\theta^i(z_m,t)=\theta_m^i(t)$.

        Differentiating the interpolation polynomial in \eqref{eq:Lagrange_temperature_approximation} with respect to $z$ 
        and evaluating the result at the collocation nodes yields
        \begin{equation} 
            \label{eq:Chebyshev_differentiation}
            \left.\frac{\partial {\theta}^{i}}{\partial z}\right|_{z = z_m} \approx \sum_{n=0}^{N-1} \theta^i_n(t) L_n'(z_m).
        \end{equation}
        Hence, the spatial derivative at each collocation node is obtained as a linear combination of all nodal temperatures. 
        This relationship defines the Chebyshev differenti\-ation matrix $\vec{D}_z \in \mathbb{R}^{N \times N}$ entrywise by
        \mbox{$\left(\vec{D}_z\right)_{mn} = L_n'(z_m)$}, with \mbox{$m,n = 0,1,\ldots, N-1$}.\newline
        Here, the row index $m$ identifies the node at which the derivative is evaluated, whereas the column index $n$ identifies the nodal temperature multiplying the corresponding basis-function derivative.

        For the ascending CL nodes defined in \eqref{eq:Chebyshev_Lobatto_nodes}, the derivatives of the Lagrange basis polynomials can be evaluated
        explicitly. 
        On the one hand, the off-diagonal entries of the differentiation matrix are given by 
        \begin{equation}
            \label{eq:Differentation_Matrix_off_diagonal}
            (\vec{D}_z)_{mn} =\frac{c_m}{c_n}\frac{(-1)^{m+n}}{z_m-z_n}, \qquad m\neq n,
        \end{equation} where $c_0=c_{N-1}=2$ and $c_n=1$ otherwise. 
        On the other hand, the diagonal entries follow from the zero-row-sum property, $(\vec{D}_z)_{mm}=-\sum_{n\neq m}(\vec{D}_z)_{mn}$. 
        This approximation of the spatial derivative at the nodes can be written in vector form as:
        \begin{equation}
            \begin{bmatrix}
                \left.\frac{\partial \theta^{i}}{\partial z}\right|_{z = z_0}, \ldots, \left.\frac{\partial \theta^{i}}{\partial z}\right|_{z = z_{N-1}}
            \end{bmatrix}^T\approx \vec{D}_z \gvec{\theta}^i.
        \end{equation}
        Since $z = 2 \zeta-1$, application of the chain rule yields \mbox{$\partial z/\partial \zeta=2$}. Hence, the differentiation matrix with respect to the normalized coordinate $\zeta$ is $\vec{D}_ \zeta=2\vec{D}_z$. 
        To formulate the system model in compact form, let $\vec{e}_n = \begin{bmatrix} 0 \ldots  0 & 1 & 0 \ldots 0 \end{bmatrix}^T \in \mathbb{R}^N$, $n = 0,1,\ldots, N-1$
        \begin{subequations}
            \label{eq:Chebyshev_semidiscrete_model}
            denote the $n$-th canonical unit vector and the embedding matrices $\vec{E}_I = \begin{bmatrix} \vec{e}_1 & \vec{e}_2 & \ldots & \vec{e}_{N-1} \end{bmatrix} \in \mathbb{R}^{N \times (N-1)}$, $\vec{E}_{II} = \begin{bmatrix}\vec{e}_0 & \vec{e}_1& \ldots & \vec{e}_{N-2} \end{bmatrix} \in \mathbb{R}^{N \times (N-1)}$.
            The reduced nodal state vectors are defined as
            \begin{equation}
                \label{eq:reduced_State_Vectors_chebyshev}
                \vec{x}^{I} = \begin{bmatrix}
                    \theta^{I}_1 & \ldots & \theta^{I}_{N-1}
                \end{bmatrix}^T, \quad  \vec{x}^{II} =\begin{bmatrix}
                    \theta^{II}_0 & \ldots & \theta^{II}_{N-2}
                \end{bmatrix}^T,
            \end{equation}
            and the full nodal temperature vectors including the boundary conditions can then be reconstructed as
            \begin{equation}
                \label{eq:full_nodal_temperature_vector_I}
                \gvec{\theta}^{I} = \vec{E}_I \vec{x}^{I} + \vec{e}_0 \theta^{I}_{\mathrm{in}}, \quad \gvec{\theta}^{II} = \vec{E}_{II} \vec{x}^{II} + \vec{e}_{N-1} \theta^{II}_{\mathrm{in}}.        
            \end{equation}
            With these definitions, the overall system can be formulated as
            \begin{align}
                \label{eq:semidiscrete_model_ODEs}
                \frac{d\vec{x}^{I}}{dt} &= -\frac{v_I}{L}\vec{E}^T_{I}\vec{D}_ \zeta\gvec{\theta}^I -\alpha^{I} U \vec{E}^T_{I}\left(\gvec{\theta}^I-\gvec{\theta}^{II}\right),\\
                \frac{d\vec{x}^{II}}{dt} &= -\frac{v_{II}}{L}\vec{E}^T_{II}\vec{D}_ \zeta\gvec{\theta}^{II} -\alpha^{II} U \vec{E}^T_{II}\left(\gvec{\theta}^{II}-\gvec{\theta}^I\right), 
            \end{align}
            and the approximated outlet temperatures are given by $\theta^{I}_{\mathrm{out}} = \vec{e}^T_{N-1} \gvec{\theta}^{I}$, \mbox{$\theta^{II}_{\mathrm{out}} = \vec{e}^T_{0} \gvec{\theta}^{II}$}.
        \end{subequations}
        Thus, the distributed-parameter model is transformed into a finite-dimensional system of ordinary differential equations while retaining the nonlinear dependence of $U$ on the flow rates.
        \subsection{Pure Transport Analysis}
        \label{ss:Transport_analysis_chebyshev}
        To characterize the transport approximation independently of the thermal coupling, the heat-transfer coefficient is again set to zero, $U=0$.
        The two flow paths are then decoupled. For $N=3$ CL nodes, the transfer function from $\theta^{I}_{\mathrm{in}}$ to $\theta^{I}_{\mathrm{out}}$ is 
        \begin{equation} 
            \label{eq:Dead_Time_Approximation_Chebyshev} 
            G_I(s) = \frac{1-\frac{\tau_I}{4}s} {1+\frac{3\tau_I}{4}s+\frac{\tau_I^2}{4}s^2},
        \end{equation}
        where $\tau_I=V_I/|q_I|$ denotes the time constant of side~$I$. 
        For three CL nodes, the collocation model has relative degree one and no direct inlet-to-outlet feedthrough. 
        Its transfer function nevertheless contains a right-half-plane zero at $s_2 = \frac{4}{\tau_I}$ and therefore remains nonminimum phase.

   \section{Results and Comparison}
        \label{sc:Results and Comparison}
        All simulations use the parameterization of the compact plate heat exchanger validated in [4], [5]. The proposed models are compared with the baseline power-balance model (PB) and a method-of-lines reference model (MOL) with 150 spatial elements.
        Two experiments are considered: an \emph{inlet-temperature step} under low-flow conditions and a \emph{step of the hot-side flow rate}. All models are initialized at the same stationary operating point.
        The primary comparison uses the same number of dynamic states. The first-spatial-moment model (FSM) employs $M=3$ elements and therefore has 12 states.
        The CL model employs $N=7$ nodes and has 12 states, whereas the baseline power-balance model employs $M=6$ elements and also has 12 states.
        A secondary comparison uses the same number of spatial subintervals; this comparison does not correspond to equal model order.
        \subsection{Inlet-Temperature Step} 
        Figures~\ref{fig:temperature_step_circuit_I_same_states} and \ref{fig:temperature_step_circuit_II_same_states} show the outlet temperatures  $\theta^{I}_\mathrm{out}$ and $\theta^{II}_\mathrm{out}$ following a step of the inlet-temperature $\theta^{II}_{\mathrm{in}}$
        from $\SI{6}{\degreeCelsius}$ to $\SI{26}{\degreeCelsius}$ at $t =\SI{0}{\second}$ at constant low-flow conditions of $q_I(t) = q_{II}(t) = \SI{5}{\liter\per\minute}$, while the inlet-temperature of  side $I$ $\theta_{\mathrm{in}}^{I}$ remains at $\SI{6}{\degreeCelsius}$.

        In contrast to the method-of-lines reference, the original power-balance approximation responds instantaneously and exhibits an initial outlet excursion in the direction of the inlet step. 
        This is due to the Padé-like structure. For an odd number of volume elements, the instantaneous outlet response occurs in the opposite direction.
        Both the first-moment approximation and the CL method eliminate the instantaneous outlet jump and more closely reproduce the delayed response of the reference model.
        The largest difference between the original and extended power-balance models occurs during the initial transport interval. During the subsequent transient, the first-moment and
        similarly the CL method response remain closer to the method-of-lines solution for both fluid circuits. At the considered model orders, both proposed approximations
        reproduce the inlet-temperature-step response more accurately than the original power-balance approximation, while maintaining stationary accuracy.
        \myFigure{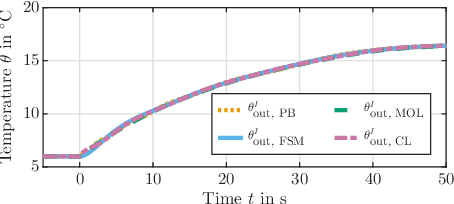}
            {width=0.96\columnwidth}
            {Outlet-temperature response of side~$I$ to a hot-side inlet-temperature step under low-flow conditions. Number of state variables:~12.}
            {blablabla}
            {temperature_step_circuit_I_same_states}
        \myFigure{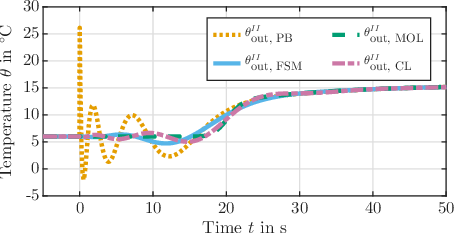}
            {width=0.96\columnwidth}
            {Outlet-temperature response of side~$II$ to a hot-side inlet-temperature step under low-flow conditions. Number of state variables:~12.}
            {blablabla}
            {temperature_step_circuit_II_same_states}
            
        By a comparison of the methods using not the same number of state variables, but the same number of spatial elements, see \autoref{fig:temperature_step_circuit_II_same_volumes}, the results of a temperature step 
        under low-flow conditions are even more in favor of the proposed approximations. In the case of three volume elements, the immediate step of the power-balance method in the opposite direction mentioned above can be observed.
        \myFigure{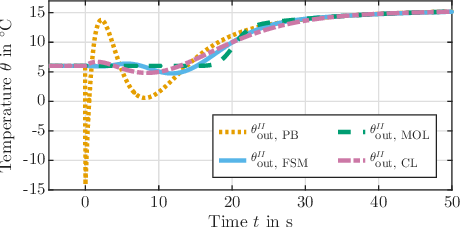}
            {width=0.96\columnwidth}
            {Outlet-temperature response of side~$II$ to a hot-side inlet-temperature step under low-flow conditions. Number of spatial elements:~3.}
            {blablabla}
            {temperature_step_circuit_II_same_volumes}
            
        \subsection{Flow-Rate Step}
        Figure~\ref{fig:flowstep_both_sides_same_states} shows the outlet temperatures $\theta^{I}_\mathrm{out}$ and $\theta^{II}_\mathrm{out}$ for a step change of the hot-side flow rate $q_{II}$, which is increased from  $\SI{5}{\litre\per\minute}$ to $\SI{25}{\litre\per\minute}$ at $t = \SI{0}{\second}$
        and returned to $\SI{5}{\litre\per\minute}$ at $t = \SI{25}{\second}$,
        while $q_I = \SI{30}{\litre\per\minute}$  and both inlet temperatures at $\theta^{I}_\mathrm{in} = \SI{6}{\degreeCelsius}$ and $\theta^{II}_\mathrm{in} = \SI{25}{\degreeCelsius}$ remain constant.
        For this experiment, all reduced order methods have the same state dimension.

        The variation affects both the convective velocity $v_{i}$ and the flow-dependent overall heat transfer coefficient $U$. For both circuits, the PB method, the CL method, and the first-moment approximation
        reproduce the dominant response of the MOL reference. The two proposed approaches remain close to the reference throughout the transient, although the differences
        between the two reduced models are small for this excitation.
        Thus, all considered low-order models provide a similar representation of the investigated flow-rate transient.

        \myFigure{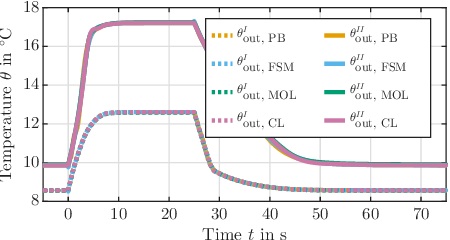}
            {width=0.96\columnwidth}
            {Outlet-temperature response of both sides to a hot-side flow-rate step. Number of state variables:~12.}
            {blablabla}
            {flowstep_both_sides_same_states}

    \section{Conclusion}
        \label{sc:Conclusion}
        The proposed first-spatial-moment (FSM) and Chebyshev--Lobatto (CL) approximations improve the low-order representation of the temperature modeling of a heat exchanger. 
        While all investigated models show similar behavior for the considered flow-rate step, the differences become more pronounced for inlet-temperature changes under low-flow conditions. 
        At \emph{equal state dimension}, both proposed methods better capture the delayed outlet response of the method-of-lines reference and substantially reduce the initial oscillations of the baseline power-balance model. 

        Consequently, the proposed models provide a more suitable trade-off between dynamic accuracy and model order when the fluids transport delay significantly affects the system behavior. 
        The proposed models are therefore promising candidates for model-based control and state estimation.
    \section*{CRediT authorship contribution statement}
    \textbf{Noah Hofmann}: Writing -- original draft, Writing -- review and editing, Conceptualization, Formal analysis, Methodology, Investigation, Software, Visualization. 
    
    \textbf{Stefan L. H\"olzl}: Writing -- review and editing, Conceptualization. 
    
    \textbf{Robert Bauer}: Writing -- review and editing, Supervision. 
    
    \textbf{Martin Steinberger}: Writing -- review and editing, Funding acquisition, Supervision.
    \section*{Declaration of generative AI and AI-assisted technologies in the manuscript preparation process}
    During the preparation of this work the authors used ChatGPT 5.6 SOL with ChatGPT Images 2.5 (OpenAI) to generate a schematic figure and for language refinement. After using this tool/service, the authors reviewed and edited the content as needed and take full responsibility for the content of the published article.
    \printbibliography
\end{document}